\documentstyle[preprint,aps]{revtex}
\tightenlines
\begin{document}

\title{Dynamical Localization of Gravity}

\author{\normalsize{Aharon Davidson\footnote{Email 
address: davidson@bgumail.bgu.ac.il}} \\
\normalsize{Physics Department,
Ben Gurion University of the Negev, Beer-Sheva 84105, Israel} \\
\normalsize{and} \\
\normalsize{Philip D. Mannheim\footnote{Email address: 
mannheim@uconnvm.uconn.edu}} \\
\normalsize{Center for Theoretical Physics, Laboratory for Nuclear
Science and Department of Physics,
Massachusetts Institute of Technology, Cambridge, Massachusetts 02139} \\
\normalsize{and} \\
\normalsize{Department of Physics,
University of Connecticut, Storrs, CT 06269\footnote{permanent address}} \\
\normalsize{(MIT-CTP-3015, hep-th/0009064 v2, October 23, 2000)} \\}

\maketitle

\begin{abstract}
We show that the thin wall limit of the thick domain wall associated with 
a sine-Gordon soliton in a single non-compactified patch of 5-dimensional
spacetime explicitly yields the Randall-Sundrum localized gravity two patch
brane, with its discrete $Z_2$ symmetry arising from the discrete symmetry 
of the potential, and with the thin Minkowski brane
$\Lambda_5+\kappa^2_5\Lambda^2_b/6=0$ relation between bulk and 
brane cosmological constants arising naturally without any need for fine 
tuning. Additionally we show that for an embedded thin de Sitter brane, 
localization of gravity is again possible provided the 5-space is 
compactified, with the now non-zero net cosmological constant
$\Lambda_5+\kappa^2_5\Lambda^2_b/6$ on the brane being found to vary 
inversely with the compactification radius.   
\end{abstract}

\bigskip

Recently Randall and Sundrum \cite{Randall1999a,Randall1999b} have shown 
that it is possible for our 4-dimensional universe to be a brane embedded in 
a 5-dimensional bulk spacetime whose spacelike extra dimension need 
not in fact be tiny. Intrinsic to their study was a $Z_2$ symmetry associated
with the piecing together of two separate patches of $AdS_5$ spacetime at the
brane in a way which then led to exponential suppression of gravity on both 
sides of the brane. It is thus of interest to find an origin for this 
otherwise simply presupposed $Z_2$ symmetry, and in this paper we will show
that both it and the emergence of the two separate $AdS_5$ patches can arise
naturally from the thin wall limit of the topological domain wall which is
generated by a gravitationally supported soliton mode associated with
a sine-Gordon scalar field $\phi$ in a curved 5-dimensional spacetime
background.

In order to eventually make contact with the $R_4$ Minkowski thin brane case
studied by Randall and Sundrum (RS) as well as try to generalize their result
to de Sitter and anti de Sitter branes, we shall consider the embedding of an
arbitrary maximally symmetric 4-space (viz. one not yet a brane) into an
arbitrary (viz. one not yet $AdS_5$) 5-space, with the most general metrics
for such embeddings of the three possible 4-spaces being given by 
\begin{eqnarray}
ds^2(R4)=dw^2+e^{2f}[dx^2+dy^2+dz^2-dt^2],
\nonumber \\
ds^2(dS4)=dw^2+e^{2f}[e^{2at}(dx^2+dy^2+dz^2)-dt^2],
\nonumber \\
ds^2(AdS4)=dw^2+e^{2f}[dx^2+e^{2bx}(dy^2+dz^2-dt^2)],
\label{1}
\end{eqnarray}
where in each case $f(w)$ is an arbitrary function of the extra coordinate
$w$, one which need not even be a symmetric function of $w$, let be alone be
one of the $f=-|w|$ form required in the RS $R_4$ Minkowski case. For the
5-space we shall take the gravitational equations of motion to be of the form
\begin{equation}
G_{AB}=R_{AB}-g_{AB}R^C_{\phantom{C}C}/2=-\kappa^2_5 T_{AB} 
\label{2}
\end{equation}
where $T_{AB}$ ($A,B=0,1,2,3,5$) is due only to a bulk scalar field $\phi$ 
with yet to be specified $\phi \rightarrow -\phi$ invariant potential
$V(\phi)$, so that the field equations then take the form:
\begin{eqnarray}
3f^{\prime \prime}+6f^{\prime 2}-3\sigma e^{-2f}=
-\kappa^2_5 e^{-2f}T_{00}=-\kappa^2_5[\phi^{\prime
2}/2+V(\phi)],
\nonumber \\
6f^{\prime 2}-6\sigma e^{-2f}=\kappa^2_5T_{55}=\kappa^2_5[\phi^{\prime
2}/2-V(\phi)],
\nonumber \\
\phi^{\prime \prime}+4f^{\prime }\phi^{\prime}=dV(\phi)/d\phi,
\label{3}
\end{eqnarray}
where $\sigma=0,a^2,-b^2$ in the three cases under discussion. As regards 
these equations we note immediately that independent of the explicit structure
of the potential, as long as the theory supports a kink or soliton which
interpolates between different degenerate vacua of $V(\phi)$, $\phi(w)$ will
then be an odd function of $w$, thus causing $f(w)$ to be an even function
of $w$. A $Z_2$ type symmetry of a potential thus translates directly into a
$Z_2$ symmetry of the geometry. Spontaneous symmetry breaking thus naturally
yields the RS $Z_2$.  

Recalling the geometric connection between maximally symmetric spaces and the
sine-Gordon  equation,\footnote{The general (Euclidean and Minkowski)
signatured 2-dimensional metrics
$ds^2=dx^2[1+cos \theta(x,y)]\pm dy^2[1- cos \theta(x,y)]$ will be spaces of
constant 2-curvature $K$ provided (see e.g. \cite{Coleman1985})
$\theta$ obeys the conditions
$\partial^2\theta/\partial_x^2\mp \partial^2\theta/\partial_y^2=-2Ksin \theta$,
and will support the $y$-independent soliton
$tan(\theta/4)=exp[(-2K)^{1/2}x]$ if $K$ is negative.} and recalling that our
sought $AdS_5$ is itself a space of constant negative 
curvature, it is thus suggested to try as an explicit potential
\begin{equation}
V(\phi)=A^2\beta^2/8-(A^2\beta^2/8)(1+\kappa^5_2A^2/3)sin^2(2\phi/A).
\label{6}
\end{equation}
On setting
\begin{equation}
tan(\phi/A)=tanh(\beta w/2),~~\phi^{\prime}=(A\beta/2)sech(\beta
w)=(A\beta/2) cos (2\phi/A),
\label{7}
\end{equation}
viz. a solitonic thick domain wall mode, we find that an exact solution to the
entire set of field equations of Eq. (3) then obtains\footnote{This solution 
was also reported in \cite{Gremm2000a,Behrndt2000} where some 
interesting implications for brane localized gravity were presented.} in the
$\sigma=0$ $R_4$ case provided ($D$ is another constant)
\begin{equation}
f^{\prime}=-(A^2\beta\kappa^2_5/12) tanh(\beta w),~~
e^f=D[cosh(\beta w)]^{-A^2\kappa^2_5/12},
\label{9}
\end{equation}
a solution we recognize as precisely being of a thick brane form.
While not being an $AdS_5$ metric,\footnote{With a spatially varying scalar 
field yielding an energy-momentum tensor which is not purely a 5-dimensional
cosmological constant, the bulk geometry could not be $AdS_5$.} $e^f$ is 
nonetheless seen to peak at $w=0$ and to fall exponentially to zero at 
$w=\pm \infty$, to thus rapidly localize the geometry to the $w=0$ region,
without any need for the domain wall to be thin. Since fluctuations of the 
form  $ds^2=dw^2+e^{2f}dx^{\mu}dx^{\nu}(\eta_{\mu\nu}+
\psi_{\mu \nu})$ around this metric obey 
\cite{Randall1999a,Randall1999b,Giddings2000,DeWolfe2000} the
covariant 5-space scalar wave equation associated with Eq. (\ref{1}), viz.
\begin{equation}
(\partial_w^2+4f^{\prime}\partial_w)\psi_{\mu \nu}
=-e^{-2f}(\partial^2_x+\partial^2_y+\partial^2_z-\partial^2_t)\psi_{\mu\nu}=
-e^{-2f}m^2\psi_{\mu \nu},
\label{10}
\end{equation}
where $m^2$ is a separation constant, we see that no matter what the explicit
form of $f(w)$, the theory will admit of an
$m^2=0$ massless 4-dimensional graviton mode with an associated metric
fluctuation $h_{\mu \nu}(w)=e^{2f} \psi_{\mu \nu}$ which behaves 
as $e^{2f}$. Localization of the geometry to the $w=0$ region thus entails
localization of gravity to the $w=0$ region as well, with $f(w)$ thus not
needing to be restricted to the particular $e^{-|w|}$ $AdS_5$ RS fall-off in
order to localize, with Eq. (\ref{9}) providing a far more general such
option.\footnote{In fact, any metric which can be written in the generic
separable $ds^2=dw^2+e^{2f(w)}dx^{\mu}dx^{\nu}q_{\mu \nu}(x,y,z,t)$ form will 
admit of a separable analog of Eq. (\ref{10}) and thus necessarily support an
$m^2=0$ mode which will move on the light cone associated with the 4-space
metric $q_{\mu \nu}(x,y,z,t)$. It is only the relative
importance of the $m^2> 0$ modes, or the possible existence of tachyons,
which would be sensitive to the explicit form of $q_{\mu \nu}$.}

As far as gravity alone is concerned, a macroscopically sized (though not too
thick) brane could actually be phenomenologically acceptable. However, in such
a non delta function thin case the $SU(3)\times SU(2) \times SU(1)$ particle
physics fields could then potentially leak into the fifth dimension as well,
with it being particle physics which thus obliges us to look for a thin brane
limit of our above model. Moreover, it is important to stress that the
localization associated with the metric of Eq. (\ref{9}) has been obtained
above with the use of only one single 5-space coordinate patch which embraces
both positive and negative $w$, whereas the particle physics required RS thin
brane entails the presence of two. It is thus very interesting to now note
that our thick brane solution can precisely be brought to the RS thin brane two
patch form by taking a very delicate limit in which we let $A$ go to zero and
$\beta$ go to infinity while holding
$A^2\beta$ fixed. Noting that in the $\beta \rightarrow \infty$ limit the 
quantity $[cosh(\beta w)]^{-1/\beta}\rightarrow e^{-|w|}$, we thus see that
$f\rightarrow -\kappa^5_2A^2\beta|w|/12$, i.e. precisely to the standard RS 
form. With the solitonic $\phi$ becoming a step function in this same limit, 
we see that we generate a thin domain wall in the limit, with this wall 
becoming a brane which then divides our original one coordinate patch into 
two now disconnected regions across which there is a discontinuous  
jump in the extrinsic curvature $K_{\mu \nu}=\eta_{\mu\nu}f^{\prime}$ 
of the brane. The solitonic mode thus generates the RS thin brane 
dynamically,\footnote{For an earlier attempt to generate a brane using the
$\phi=Atanh(\beta w)$ kink mode  see 
\cite{DeWolfe2000}.} to not only recover it, but to also put it on a far more 
secure theoretical foundation, one based on structures quite familiar in 
particle physics (the higher dimensional $\phi$ itself might even be related
to a string theory dilaton). 

On identifying the value of the sine-Gordon potential at its degenerate 
minima as $V(\pm \pi A/4)=\Lambda_5=-\kappa^2_5A^4\beta^2/24$, we thus see 
that $f \rightarrow -(-\Lambda_5\kappa^2_5/6)^{1/2}|w|$ in the brane limit,
so that $f^{\prime} \rightarrow -(-\Lambda_5\kappa^2_5/6)^{1/2}\epsilon(w)$
(here $\epsilon(w)=\theta(w)-\theta(-w)$), while $f^{\prime\prime}\rightarrow 
-2(-\Lambda_5\kappa^2_5/6)^{1/2}\delta(w)$, to thus generate jump and delta
function singularities at the brane. In order to determine exactly where this 
delta function comes from in our model, we note from Eq. (\ref{3}) that the
simultaneous presence of a delta function singularity in $f^{\prime\prime}$ 
and the absence of one in $f^{\prime}$ entail that $\phi^{\prime 2}/2$ and 
$V(\phi)$ must both contain the same delta function. And indeed from Eq.
(\ref{7}) we see immediately that  $\phi^{\prime 2}\rightarrow 
A^2\beta \delta(w)/2$ in the limit,
to thus generate a term $e^{2f}[\phi^{\prime 2}/2+V(\phi)]
\rightarrow (A^2\beta/2)\delta(w)$ in $T_{00}$, an effective brane
cosmological constant  term $\Lambda_b\delta(w)=(A^2\beta/2)\delta(w)$ which
then automatically obeys the condition $\Lambda_5+\kappa^2_5\Lambda_b^2/6=0$.
Thus what must be thought of as a fine tuning condition in an RS theory with
an a priori brane possessing a $\Lambda_b$  which is introduced by hand is now
seen as being a dynamical output to a soliton induced brane, with $\Lambda_b$
being due to the energy density of the bulk soliton when it is squeezed to
the selfsame brane which it itself generates.

It is also instructive to derive this effective $\Lambda_b$ in a slightly
different manner. Specifically, we can also define it globally, with the 
action $-\int dw g^{1/2}L= \int dw e^{4f}[\phi^{\prime2}/2+V(\phi)]
=-3\int dw e^{4f}(f^{\prime\prime}+2f^{\prime 2})/\kappa^2_5$ being required to
recover the RS action $\int dw e^{2f}[e^{2f}\Lambda_5+\Lambda_b\delta(w)]$ in 
the thin brane limit. Explicit calculation then shows that this is indeed the
case while revealing that the
$\Lambda_b$ term comes entirely from the $f^{\prime\prime}$
term according to $\Lambda_b=A^2\beta/2$, with the $\Lambda_5$ term coming
entirely from the $f^{\prime 2}$ term according to  
$\Lambda_5=-A^4\beta^2\kappa^2_5/24=-\kappa^2_5\Lambda_b^2/6$.
Moreover, not only does $\Lambda_b$ come purely from the $f^{\prime\prime}$
integral, in the limit  the integral evaluates to
$\Lambda_b=-3[f^{\prime}(+\infty)-f^{\prime}(-\infty)]/\kappa^2_5$. With
$f^{\prime}$ being given by $-(\kappa^2_5A^2\beta/12)tanh(\beta w) 
\rightarrow-(\kappa^2_5A^2\beta/12)\epsilon(w)$, we may thus interpret the
induced $\Lambda_b=A^2\beta/2$ as a topological 
charge.\footnote{Since Eq. (1) describes the general embedding of a 4-space into
a  5-space, already before taking any thin brane limit we may set 
$K_{\mu\nu}(\infty)-K_{\mu\nu}(-\infty)=
\eta_{\mu \nu}[f^{\prime}(\infty)-f^{\prime}(-\infty)]$. Since we then obtain 
$f^{\prime}(\pm\infty)=f^{\prime}(0^{\pm})$ in the ensuing thin brane limit, we
may thus write  $K_{\mu\nu}(\infty)-K_{\mu\nu}(-\infty)=
\eta_{\mu \nu}[f^{\prime}(0^{+}) -f^{\prime}(0^{-})]=
K_{\mu\nu}(0^{+})-K_{\mu\nu}(0^{-})$. Then, since the discontinuity at the brane
obeys the Israel junction condition $K_{\mu\nu}(0^{+})-K_{\mu\nu}(0^{-})=
-\kappa^2_5(T_{\mu \nu}-q_{\mu
\nu}T^{\alpha}_{\alpha}/3)=- \kappa^2_5\eta_{\mu \nu}\Lambda_b/3$, we can thus,
and with full generality, identify $\Lambda_b$ as the topological
$f^{\prime}(\infty)-f^{\prime}(-\infty)=-\kappa^2_5\Lambda_b/3$. As a measure 
of the universality of such a topological charge, we note in passing that the
sextic potential 
$V(\phi)=A^2\beta^2(1-\phi^2/A^2)^2/2-2\kappa^2_5A^2\beta^2\phi^2
(3-\phi^2/A^2)^2/27$ (a potential also considered in \cite{DeWolfe2000})
admits of the exact solution $\phi=Atanh(\beta w)$, 
$e^f=exp[A^2\kappa^2_5/18cosh^2(\beta w)]
cosh(\beta w)^{-2A^2\kappa^2_5/9}$, a solution which also possesses an RS
brane limit in which
$\Lambda_b=(-6V_{min}/\kappa^2_5)^{1/2}=
-3[f^{\prime}(+\infty)-f^{\prime}(-\infty)]/\kappa^2_5$.} 

Turning now to the $dS_4$ and $AdS_4$ embedded cases, we note first that 
while the general 5-dimensional metrics of Eqs. (\ref{1}) are not
necessarily $AdS_5$ metrics, requiring their associated Einstein
tensors to be given as their respective metric tensors\footnote{With the Weyl
tensor vanishing identically for arbitrary $f(w)$ for each of the three
metrics in Eqs. (\ref{1}), setting each Ricci tensor equal to the
relevant metric tensor makes each such metric maximally 5-space 
symmetric.} would make them so. For $R_4$ this leads to 
$e^f=e^{\pm w}$, for $dS_4$ to $e^f=sinhw$ and for $AdS_4$ to
$e^f=coshw$. With both of the $dS_4$ and $AdS_4$ metrics thus
growing at both large positive and large negative $w$, prospects for
localization of gravity in these cases initially appear to be somewhat 
slim.\footnote{Exactly this same situation is encountered in the 
embedding of maximally 3-symmetric Robertson-Walker branes in a general
5-space \cite{Mannheim2000a} or even into an $AdS_5$ one
\cite{Mannheim2000b}.} In order to see whether these cases could anyway
support a brane (regardless of any possible localization of gravity
considerations), the  natural thing to do is consider metrics of the forms
$e^f=sinh|w|$ and  $e^f=cosh|w|$. However, neither of these cases works as
such since even while their Einstein tensors now contain $\delta(w)$ terms,
the coefficients of these delta functions respectively diverge and vanish at 
$w=0$. However, it is possible to avoid these difficulties by simply
shifting the position of these singularities elsewhere, to thus instead take
$e^f=sinh(\gamma-|w|)/sinh\gamma$ and $e^f=cosh(\gamma-|w|)/cosh\gamma$
where $\gamma$ is a constant.\footnote{The thin brane
$e^f=sinh(\gamma-|w|)$ metric was also considered in
\cite{Kim2000,DeWolfe2000,Tye2000}, with discussion of a thick $dS_4$ brane
being given in \cite{Gremm2000b}.}
While the choice of sign of the shift parameter
$\gamma$ is arbitrary, we note that for positive $\gamma$ both of the metrics
would then have a local maximum at $w=0$. Thus, if the metrics could be
cut-off at
$w=\pm \gamma$ by compactifying the fifth dimension into a circle via an
identification of $w=+\gamma$ with $w=-\gamma$, gravity would then fall off
on both sides of the brane.\footnote{Since the geometry would fall off 
exponentially on both sides of the brane for $|w| \ll \gamma$, and
since both the sinh and cosh  are monotonic functions of their arguments, the
geometry would then fall off all the way to $w=\pm \gamma$, so that the
compactification radius $\gamma$ need not be as tiny as required in standard
compactified empty bulk Kaluza-Klein theories in which the 5-dimensional
geometry is taken to be the compactified but otherwise flat $S_1\times R_4$.}
With the $dS_4$ metric actually vanishing at the compactification radius, and
with derivatives of $f(w)$ diverging there, the very existence of such 
singular behavior could cause the $|w|>\gamma$ region to be cut-off, and thus 
provide a mechanism for compactification in the first place, to thus make 
$dS_4$ a potentially promising candidate for a possible such
self compactification.

While we will explicitly explore this possibility in detail below, before
doing so however, we find it instructive to first consider the
$e^f=sinh(\gamma-|w|)/sinh\gamma$ and $e^f=cosh(\gamma-|w|)/cosh\gamma$
cases in their own right simply as RS type thin brane theories with given a
priori $T_{00}=e^{2f}\Lambda_5+\Lambda_b \delta(w)$, $T_{55}=-\Lambda_5$
cosmological constant sources. In such  a situation both of these metrics are
then found to be exact RS type solutions to the Einstein equations, with their
various parameters being related according to 
\begin{equation}
a^2sinh^2\gamma=-\kappa^2_5\Lambda_5/6=1,~~
\Lambda_5+\kappa_5^2\Lambda_b^2/6=6a^2/\kappa^2_5 >0
\label{11}
\end{equation}
in the $dS_4$ case, and to
\begin{equation}
b^2cosh^2\gamma=-\kappa^2_5\Lambda_5/6=1,~~
\Lambda_5+\kappa_5^2\Lambda_b^2/6=-6b^2/\kappa^2_5 <0 
\label{12}
\end{equation}
in the $AdS_4$
one. From the point of view of pure RS theory then, we see that when the
brane $\Lambda_b$ does not obey the $\Lambda_5+\kappa_5^2\Lambda_b^2/6=0$
$R_4$ fine tuning condition, the brane must instead be $dS_4$ or 
$AdS_4$ according to whether the excess or residual brane cosmological
constant $\Lambda_b-(-6\Lambda_5\kappa^2_5)^{1/2}$ is positive or
negative.\footnote{With the fine-tuned $R_4$ RS value for $\Lambda_b$ arising
from a higher dimensional soliton, this excess brane cosmological constant
could  be thought of as being due to the contribution of the ordinary
4-dimensional particle physics fields on the brane; and with these latter
fields also obeying curved space wave equations such as the equation 
$[\partial_w^2+4f^{\prime}\partial_w+e^{-2f}\nabla^2)]\psi=0$ discussed
above, we thus note that these latter fields will also be confined to the
brane by the 5-dimensional soliton.} Thus no matter what the value of
$\Lambda_b$, there will always be an appropriate associated maximally
4-symmetric brane, so that the RS fine-tuning condition does not in
fact need to be imposed even in theories with a priori branes. Rather,
the brane topology adjusts itself according to whatever value $\Lambda_b$ 
has. Moreover, since $6a^2/\kappa^2_5=6/\kappa^2_5sinh^2\gamma$, we see 
in the event of any such excess, that the residual brane cosmological
constant would then vary inversely with the compactification
radius.\footnote{A similar result was also noted in \cite{Tye2000} in a two
brane model in which the branes were located at the $w=\pm \gamma$
compactification points.}  Consequently, a large compactification radius
(something now possible with localized gravity) entails a small residual
brane cosmological  constant. While this is a very interesting such
correlation, it is important to note that this does not yet constitute a
solution to the cosmological constant problem until some independent reason
is identified which would lead to a large compactification radius in the
first place, since a large particle physics scale cosmological constant on
the brane would itself otherwise lead only to a small compactification
radius.\footnote{For a completely different approach to the cosmological
constant problem, a strictly 4-dimensional one in which the cosmological 
constant can have acceptably small observable consequences for 4-dimensional
physics even when being as big as particle physics suggests see
\cite{Mannheim1999}.}       

In trying to construct the above $dS_4$ brane as a limiting solution to a
5-dimensional scalar field theory, we shall try to parallel the Minkowski
case brane discussion by looking for a soliton like solution in which the
scalar field goes to separate degenerate minima, $\phi(\pm\gamma)$, at the two
compactification  points, with $\phi^{\prime}$ vanishing at those
points.\footnote{In passing we note that the field $tan \phi =2\gamma
w/(\gamma^2-w^2)$ obeys
$\phi^{\prime}=2\gamma/(\gamma^2+w^2)=(1+cos\phi)/\gamma$, and is thus
actually associated with a compactified flat space sine-Gordon equation.
However, in such a configuration $\phi^{\prime}$ is non-vanishing at the
compactification points, and thus not of interest for our purposes here.}
However now, given  the singularity structure of the
$e^f=sinh(\gamma-|w|)/sinh\gamma$, 
$f^{\prime}=-\epsilon(w)coth(\gamma-|w|)$ metric at $w=\pm \gamma$, we need
to determine how a singular behavior such as this (assuming it to be present
prior to taking the brane limit) would affect the scalar field at those points.
Thus, taking the scalar field to behave as 
$\phi(w)=\phi(\gamma)+E(\gamma-w)^n$ near $w=\gamma$, then entails that
$V^{\prime}(\phi)$ must behave as 
$\phi^{\prime\prime}+4f^{\prime}\phi^{\prime}=En(n+3)(\gamma-w)^{n-2}$
near the compactification points. If $V(\phi)$ is to be a well-behaved, Taylor
series expandable, potential with a smooth minimum at $w=\gamma$, then it must
behave near such a minimum as 
$V^{\prime \prime}(\phi(\gamma))[\phi-\phi(\gamma)]=EV^{\prime
\prime}(\phi(\gamma))[\gamma-w]^{n}$, a form which we see is simply not
compatible with $\phi^{\prime \prime}+4f^{\prime}\phi^{\prime}$ no matter what
the value of $n$. Hence, near its minima $V(\phi)$ cannot in fact be as
well-behaved as a standard particle physics Higgs potential.  
Rather, $V^{\prime}(\phi)$ must also behave as
$(\gamma-w)^{n-2}$, i.e. as $[\phi(\gamma)-\phi]^{(n-2)/n}$, so that 
the (necessarily real) potential itself  must behave as the
non-analytic $(|\phi-\phi(\gamma)|)^{2(n-1)/n}$ and thus have a cusp at 
$w=\gamma$. With the same behavior having to occur at $w=-\gamma$ as well,
rather than being a smooth double well potential, $V(\phi)$ would have to be 
a double or periodic ($\simeq sin|\phi|$) cusp potential, with the loss of
analyticity at the two cusp minima signaling the presence of the singular
compactification points, while also potentially providing us with a mechanism
to disconnect the
$|w|<\gamma$ and $|w|>\gamma$ regions (possibly even by putting additional 
branes at the compactification points).

Unfortunately, for the moment, we have so far been unable to find any
exact solutions to the theory which have this particular cusp structure
at the compactification points and then produce a brane at $w=0$ when the 
brane limit is taken. However, we have, instead, been able to find a 
different type of solution, one with another kind of singularity structure,
viz. one in which the scalar field itself also diverges at the
compactification points. And despite the fact that this is an at first
somewhat disquieting requirement, we shall now show that not only is it
feasible, but that it is even achievable without any divergence in the energy
density. In fact, we have actually found a closed form solution to the field
equations at the brane limit, one which explicitly allows a divergent $\phi$ 
to support a $dS_4$ brane. Specifically, the field configuration (viz. a
configuration for which
$\phi(w=\pm \gamma)=\pm \infty$) 
\begin{equation}
e^{2\phi/\nu}=\epsilon(w)[cosh(\gamma-|w|)+1]/[cosh(\gamma-|w|)-
1]-e^{2\mu/\nu},~e^{f}=sinh(\gamma-|w|)/sinh\gamma
\label{13}
\end{equation}
(here $e^{2\mu/\nu}=\epsilon(w)[cosh\gamma+1]/[cosh\gamma-1]$) is
found to be an exact solution to the field equations when the
potential is taken to have the "shine-Gordon" form\footnote{The general 
(Minkowski and Euclidean) signatured 2-dimensional metrics
$ds^2=dx^2[1+cosh \theta(x,y)]\pm dy^2[1- cosh \theta(x,y)]$ will be spaces of
constant 2-curvature $K$ provided $\theta$ obeys the conditions
$\partial^2\theta/\partial_x^2\mp \partial^2\theta/\partial_y^2=-2Ksinh
\theta$, and will support the $y$-independent mode  
$tanh(\theta/2)=sin[(-2K)^{1/2}x]$ if $K$ is negative, a mode
in which $\theta$ becomes infinite at finite $x$.}              
\begin{equation}
V=V_0-3\nu^2sinh^2(\phi/\nu+\mu/\nu)/2,
\label{14}
\end{equation}
and $T_{00}$ is taken to also include an explicit excess brane
cosmological constant according to $e^{-2f}T_{00}=\phi^{\prime
2}/2+V(\phi)+\Lambda_b\delta(w)$ ($T_{55}=\phi^{\prime2}/2-V(\phi)$ is taken
to be unchanged). In this solution the various parameters are  found to
obey
\begin{equation}
a^2sinh^2\gamma+\kappa^2_5\nu^2/3=1,~~
-\kappa_5^2V_0/6=1,~~
\Lambda_b^2\kappa^2_5/6+V_0=6a^2/(\kappa_5^2-\kappa_5^4\nu^2/3),
\label{15}
\end{equation}
to yield a structure very similar to that exhibited in Eq. (\ref{11}).
Despite the fact that our potential is unbounded from below, the quantity
$e^{2f}[\phi^{\prime 2}/2+V(\phi)]$ remains bounded
because of compensating zeroes in the metric. Thus while a potential such as
that of Eq. (\ref{14})  could not be considered in flat space physics, in the
event of curvature it is possible for the energy density in the gravitational
field to  compensate and make the theory well behaved. With such a diverging
of $\phi$ at the compactification points we thus uncover a mechanism
which actually serves to disconnect the $|w|<\gamma$ and $|w|>\gamma$ regions
and thereby force compactification upon us in the first place. Moreover,
we can even consider the points $\phi=\pm\infty$ as being effective "minima"
of the theory since they are the points at which $V(\phi)$ takes its lowest
values  within the compactification region; with the point $\phi=0$ at which
the brane is located being a local maximum, one about which gravity is then
localized.

While we have not been able to find a closed form generalization of this
solution beyond the brane limit, or found any solution at all that 
that would lead to Eq. (\ref{13}) in the same way that Eq. (\ref{9}) led to
the $R_4$ RS limit, nonetheless the existence of a connection between scalar
field theory and a $dS_4$ RS brane in the limit strongly suggests the
existence also of such a connection even before the brane limit is taken.  We
are indebted to Drs. M. Gremm and K. Berhndt for notifying us of their work
following the release of an earlier version of this manuscript. P.
D. Mannheim would like to thank Drs. R. L. Jaffe and A. H. Guth for the  kind
hospitality of the Center for Theoretical Physics at the Massachusetts 
Institute of Technology where part of this work was performed.  The work of P.
D. Mannheim  has been supported in part by funds provided by  the U.S.
Department of Energy (D.O.E.) under cooperative research agreement 
\#DF-FC02-94ER40818 and in part by grant \#DE-FG02-92ER40716.00.

\end{document}